\documentclass[conference]{IEEEtran}
\IEEEoverridecommandlockouts
 
\usepackage{cite}
\usepackage{amsmath,amssymb,amsfonts}
\usepackage{algorithmic}
\usepackage[ruled,norelsize]{algorithm2e}
\usepackage{graphicx}
\usepackage{textcomp}
\usepackage{xcolor}
\usepackage{hyperref}
\def\BibTeX{{\rm B\kern-.05em{\sc i\kern-.025em b}\kern-.08em
    T\kern-.1667em\lower.7ex\hbox{E}\kern-.125emX}}
    
\begin{document}

\title{DeepEMO: Deep Learning for Speech Emotion Recognition
}

\author{\IEEEauthorblockN{Enkhtogtokh Togootogtokh} 
\IEEEauthorblockA{\textit{Technidoo Solutions Lab} \\
\textit{Technidoo Solutions Germany and Mongolian University of Science and Technology}\\
Bavaria, Germany \\
enkhtogtokh.java@gmail.com, togootogtokh@technidoo.com} 
\and 
\IEEEauthorblockN{Christian Klasen}
\IEEEauthorblockA{\textit{Technidoo Solutions Lab} \\
\textit{Technidoo Solutions Germany }\\
Bavaria, Germany \\
klasen@technidoo.com}
}

\maketitle

\begin{abstract}
We proposed the industry level deep learning approach for speech emotion recognition task. In industry, carefully proposed deep transfer learning technology shows real results due to mostly low amount of training data availability, machine training cost, and specialized learning on dedicated AI tasks. The proposed speech recognition framework, called DeepEMO, consists of two main pipelines such that preprocessing to extract efficient main features and deep transfer learning model to train and recognize.  Main source code is in ~\href{https://github.com/enkhtogtokh/deepemo}{https://github.com/enkhtogtokh/deepemo} repository. 
\end{abstract}


\section{Introduction}\label{intro}
Ability to understand and manage emotions, called emotional intelligence, has been shown to play an important role in decision-making. Some researchers suggest that emotional intelligence can be learned and strengthened it refers to the ability to perceive, control, and evaluate emotions. For machine intelligence, it is also important role to understand and even generate such emotional intelligence. Here we proposed the simple yet effective speech emotional recognition modern deep learning technique called DeepEMO framework. It has been conceived considering the main AI pipeline (from sensors to results) together with modern technology trends. The DeepEMO has two main pipelines which are to extract strong speech features and deep transfer learning for the emotion recognition task. We applied them on english emotional speech case. Generally it is possible to apply them on any natural language.
There are inevitable demands to recognize the speech emotion with advanced technology.

Concretely, the key contributions of the proposed work are:

\begin{itemize}
\item The industry level AI technology for speech emotion recognition
\item The speech recognition general modern deep learning framework for similar tasks
\end{itemize}

Systematic experiments conducted on real-world acquired data have shown as: 
\begin{itemize}
\item It is possible to be common framework for many type of speech recognition task.
\item It is possible to achieve 99.9\% accuracy on well prepared training data to recognize.
\item It is possible to later generate realistic enough synthetic emotional speech data generation with multiple variations
\end{itemize}

The rest of the paper is organized as follows.
The proposed framework is described in Section \ref{proposedarch}.
The recognition deep convolutional  model is explained in Section \ref{recognition}. 
The details about the
experimental results are presented in Section \ref{experimentalresult}.
Finally, Section \ref{conclusion} provides the conclusions and future work.

\section{The proposed method (DeepEMO)}\label{proposedarch}
\begin{figure}[!htbp]
  \centering
  \includegraphics[width=\linewidth]{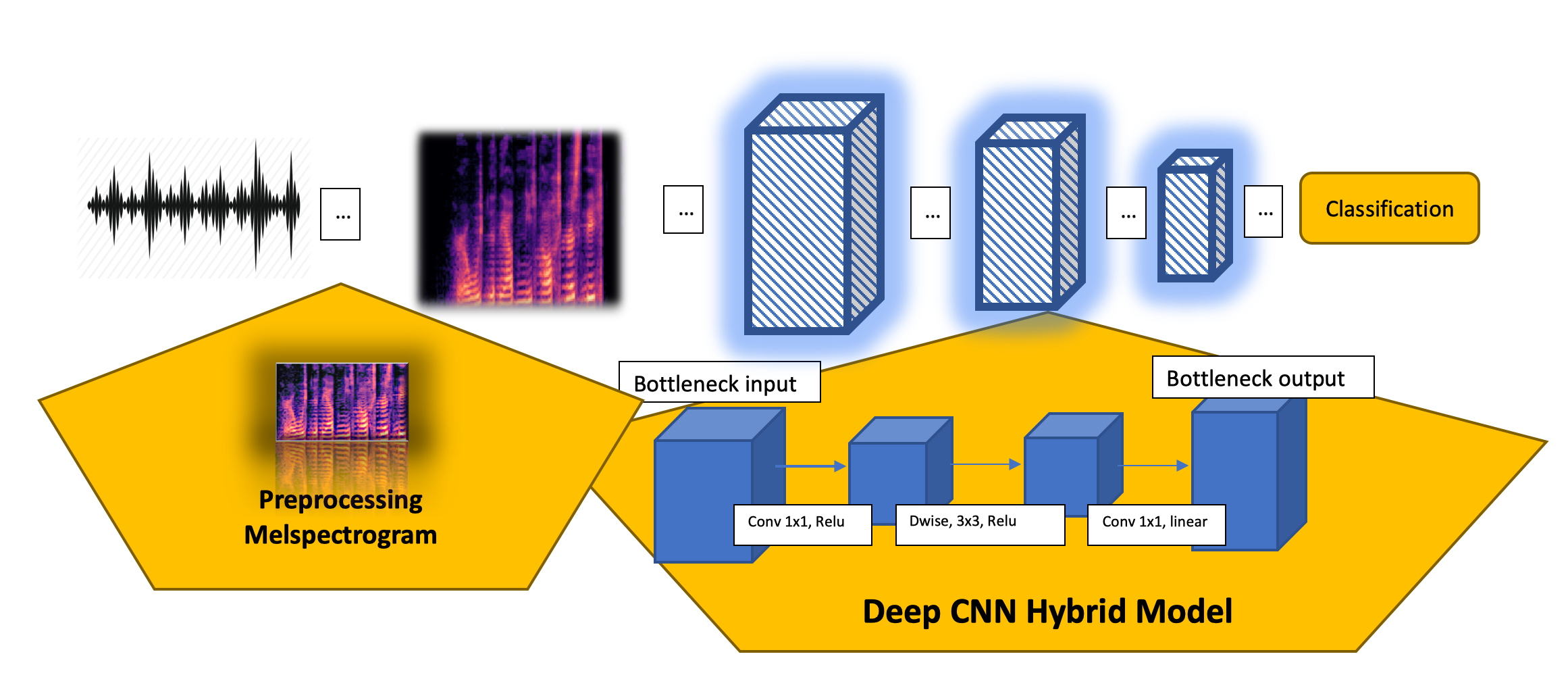}
  \caption{The DeepEMO framework}
 
  \label{fig:framework}
\end{figure}
In this section, we discuss the proposed DeepEMO model for speech emotion recognition AI applications as shown in Figure \ref{fig:framework}.
The DeepEMO has two main pipelines which are the preprocessing to extract efficient features and deep transfer learning mechanism. We discuss them in detail with coming sections.

\subsection{The preprocessing feature extraction}\label{preprocess}
Audio significant feature extraction is the important part of modern deep learning. There are many mechanisms to do it. Here we extract melspectrogram audio feature later to train machine with high accuracy.    

Specifically, it consists of following general steps:
\begin{itemize}
\item To compute fast Fourier transform (FFT)
\item To generate mel scale
\item To generate spectrogram
\end{itemize}

The FFT is an algorithm which efficiently computes the Fourier transform. The Fourier transform is a mathematical formula which decomposes a signal into it’s individual frequencies and the frequency’s amplitude. In other words, it converts the signal from the time domain into the frequency domain. The result is called a spectrum. 
The mathematical operation converts frequencies to the mel scale. Researchers proposed a unit of pitch such that equal distances in pitch sounded equally distant to the listener, which is called the mel scale. 
When signal’s frequency content varies over time as non periodic signals, it needs a right representation. As we can compute several spectrums by performing FFT on several windowed segments.  It is called the short-time Fourier transform. The FFT is computed on overlapping windowed segments of the signal, which is called the spectrogram.

\subsection{The deep transfer learning}\label{recognition}

Specifically, we define the transfer learning model by\cite{enkhtogtokh}:
\begin{itemize}
\item To prepare the pre-trained model
\item To re-define the last output layer as n (in case, n=8) neurons layer for the new task 
\item To train the network 
\end{itemize}

This is called transfer learning, i.e. we have a model trained on another task, and we need to tune it for the new dataset we have in hand.

To recognize speech emotion, we propose the deep convolutional transfer neural network. Since after melspectrogram feature extraction preprocessing, it is now generally computer vision problem.  The deep convolutional backbone model is ResNet18\cite{resnet18} which consists of assemble of convolutional layers and batch norms as shown in Algorithm \ref{alg:deepemo}. For simplicity, the Pytorch\cite{pytorch} style pseudo code is provided.
Cross Entropy Loss function (CE) and Adam optimize implemented for the model. 
\begin{algorithm}
\caption{Transfer Learning Model} \label{alg:deepemo}
\begin{algorithmic}[1]
 
\STATE import torch
\STATE import torch.nn as nn
\STATE import torchvision
\texttt{\\}
\texttt{\\}
\STATE  $model = torchvision.models.resnet18(pretrained=True)$
\STATE  $loss = torch.nn.CrossEntropyLoss()$
\STATE  $model.fc = torch.nn.Linear(in_features=512, out_features=8)$ 
\STATE  $optimizer = torch.optim.Adam(model.parameters(), lr=3e-5)$

\end{algorithmic}
\end{algorithm}

\newpage
\section{Experimental Results}\label{experimentalresult}
In this section, we discuss first about the setup, and then evaluate the deep transfer learning recognition and melspectrogram generation results are experimented in systematic scenarios. 
\subsection{Setup}
We train and test on ubuntu 18 machine with capacity of (CPU: Intel(R) Xeon(R) CPU @ 2.20GHz, RAM:16GB, GPU: NVidia GeForce GTX 1070, 16 GB). 

\subsection{The dataset}

We use the Ryerson Audio-Visual Database of Emotional Speech and Song (RAVDESS)\cite{ravdess} dataset. It has 8 speech label emotions as neutral, calm, happy, sad, angry, fearful, disgust, and surprised speeches. 

\subsection{The recognition results}
Table \ref{tab:rec} shows the accuracy of training and validation on number of epochs. After 42 epochs, we achieved enough accuracy as loss, training, and validation are 100\%, 0,009, and 100\%, correspondingly. 

\begin{table} [h!]
\centering  
\begin{tabular}{||c c c c||} 
 \hline
 Number of epoch & Training accuracy & Loss & Validation accuracy \\  
 \hline\hline
 10 & 0.88 & 0.470 & 0.970 \\ 
 \hline
 20 & 0.991 & 0.011 & 1.000 \\
 \hline
 42 & 1.000 & 0.009 & 1.000 \\
 \hline
 50 & 1.000 & 0.006 & 1.000 \\
 \hline
\end{tabular}
\caption{The deep convolutional neural network recognition model training and validation accuracy on epochs.}
\label{tab:rec}
\end{table}

\begin{figure}[!htbp]
  \centering
  \includegraphics[width=\linewidth]{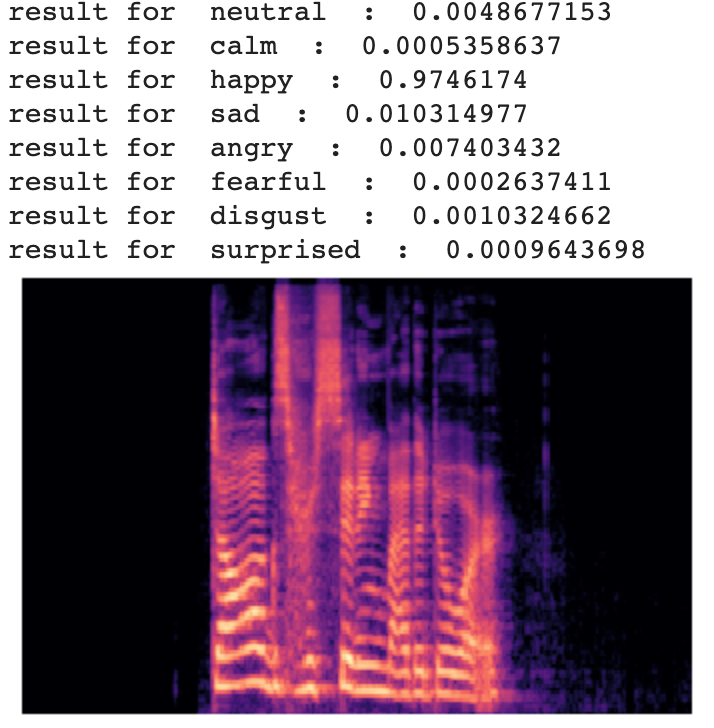}
  \caption{The recognition result for happy emotional speech}
 
  \label{fig:rec_happy}
\end{figure}

\begin{figure}[!htbp]
  \centering
  \includegraphics[width=\linewidth]{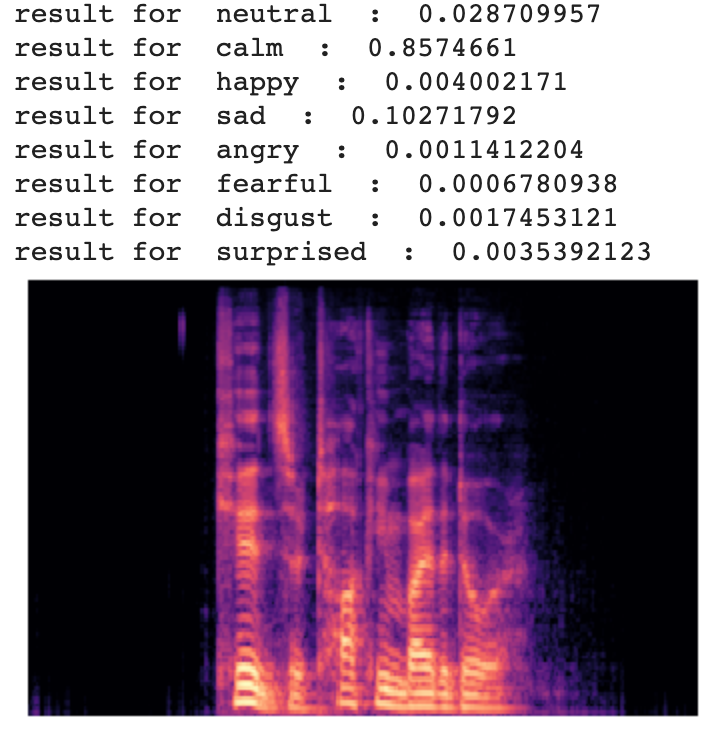}
  \caption{The recognition result for calm emotional speech}
 
  \label{fig:rec_calm}
\end{figure}

\begin{figure}[!htbp]
  \centering
  \includegraphics[width=\linewidth]{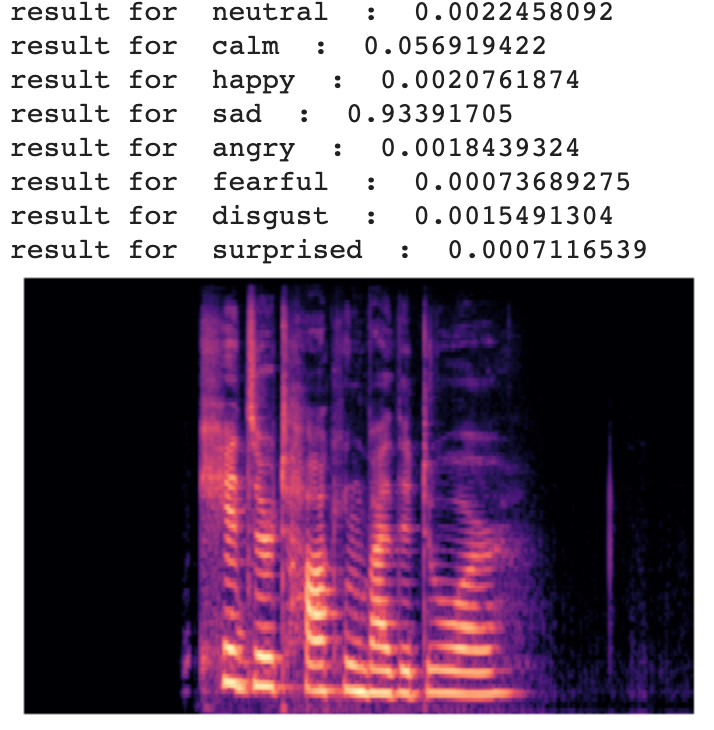}
  \caption{The recognition result for sad emotional speech}
 
  \label{fig:rec_sad}
\end{figure}

\begin{figure}[!htbp]
  \centering
  \includegraphics[width=\linewidth]{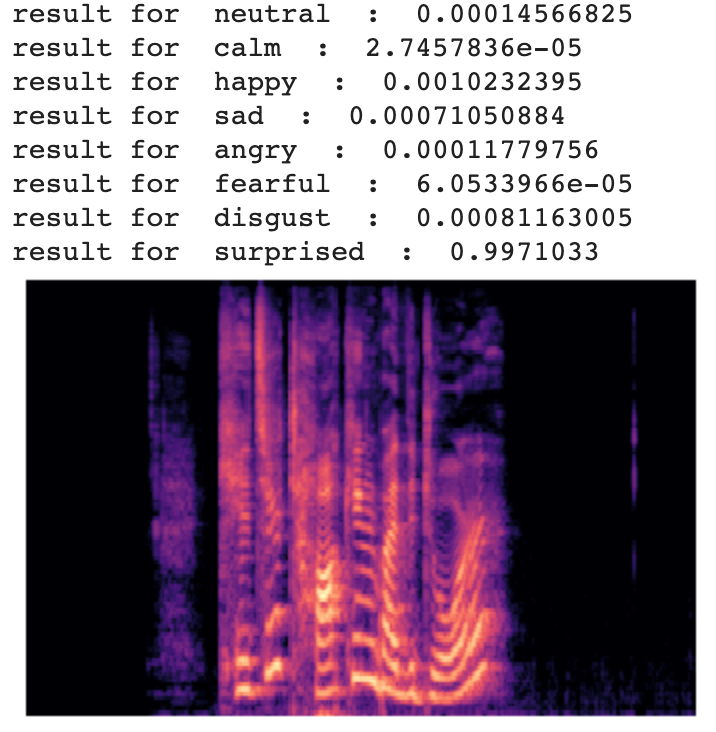}
  \caption{The recognition result for surprised emotional speech}
 
  \label{fig:rec_surprised}
\end{figure}

Figure \ref{fig:rec_happy}, \ref{fig:rec_calm}, \ref{fig:rec_sad}, and \ref{fig:rec_surprised}  show the recognition results of testing data happy, calm, sad, and surprised emotional speeches, accordingly. We printed out top-8 probability classes to show cases with corresponding melspectrogram. 
 
\newpage
\section{Conclusion}\label{conclusion}
We proposed the modern AI deep learning framework as DeepEMO for speech emotional recognition and application for industry use case. Modern state-of-the-art deep learning approaches implemented to recognize the typical emotional speeches. Main algorithm is directly provided in this research to develop first phase of emotion recognition. The real visual results and some important evaluation accuracy scores are presented. 
In future works, we will publish next series of research to apply on emotional speech generation deep learning tasks.

\vspace{12pt}

\end{document}